\documentclass{webofc}
\usepackage[varg]{txfonts} 

\usepackage{comment}

\usepackage{listings}
\usepackage[colorlinks=true, citecolor=blue, linkcolor=blue, urlcolor=blue, pdfborder={0 0 0}, breaklinks=true]{hyperref} 
\renewcommand{\url}[1]{\href{#1}{#1}}
\renewcommand{\doi}[1]{\url{https://doi.org/#1}}

\usepackage{datetime2}
 
\begin{document}

\title{\vspace*{-1.6cm}
Using HEP experiment workflows for the benchmarking\\
and accounting of WLCG computing resources}

\author{
\firstname{Andrea} \lastname{Valassi}
\inst{1}\fnsep
\thanks{\email{andrea.valassi@cern.ch} 
}
\and
\firstname{Manfred} \lastname{Alef}
\inst{2}
\and
\firstname{Jean-Michel} \lastname{Barbet}
\inst{3}
\and
\firstname{Olga} \lastname{Datskova}
\inst{1}
\and
\firstname{Riccardo} \lastname{De~Maria}
\inst{1}
\and
\firstname{Miguel} \lastname{Fontes Medeiros}
\inst{1}
\and
\firstname{Domenico} \lastname{Giordano}
\inst{1}
\and
\firstname{Costin} \lastname{Grigoras}
\inst{1}
\and
\firstname{Christopher} \lastname{Hollowell}
\inst{4}
\and
\firstname{Martina} \lastname{Javurkova}
\inst{5}
\and
\firstname{Viktor} \lastname{Khristenko}
\inst{1,6}
\and
\firstname{David} \lastname{Lange}
\inst{7}
\and
\firstname{Michele} \lastname{Michelotto}
\inst{8}
\and
\firstname{Lorenzo} \lastname{Rinaldi}
\inst{9}
\and
\firstname{\mbox{Andrea}} \lastname{Sciab\`a}
\inst{1}
\and
\firstname{Cas} \lastname{Van Der Laan}
\inst{1}
}
\institute{
CERN, Geneva, Switzerland
\and
KIT, Karlsruhe, Germany
\and
CNRS-SUBATECH, Nantes, Frances
\and
Brookhaven National Laboratory, USA
\and
University of Massachusetts Amherst, USA
\and
University of Iowa, USA
\and
Princeton University, USA
\and
INFN, Padova, Italy
\and
Universit\`a di Bologna, Italy
}

\abstract{
Benchmarking of CPU resources in WLCG 
has been based 
on the HEP-SPEC06 (HS06) suite 
for over a decade. 
It has recently become clear
that HS06,
which is based on real applications from non-HEP domains,
no longer describes typical HEP workloads.
The aim of the HEP-Benchmarks project 
is to develop 
a new benchmark suite 
for WLCG compute resources,
based on real applications from the LHC experiments.
By construction, 
these new benchmarks 
are thus guaranteed 
to have a score 
highly correlated 
to the throughputs of HEP applications,
and a CPU usage pattern 
similar to theirs.
Linux containers and the CernVM-FS filesystem
are the two main technologies enabling this approach,
which had been considered impossible in the past. 
In this paper, we review 
the motivation, implementation
and outlook
of the new benchmark suite.
}

\maketitle

\vspace*{-1mm}
\section{Introduction}
\label{sec:intro}

The Worldwide LHC Computing Grid (WLCG)
is a large distributed computing infrastructure
serving scientific research in 
High Energy Physics (HEP).
WLCG was set up to address 
the scientific computing needs 
of the four Large Hadron Collider (LHC) experiments,
and it integrates
storage and compute resources at
almost 200 sites in over 40 countries~\cite{bib:campana-espp2019}.
While 
the experiment requirements
are managed centrally 
through a well defined process~\cite{bib:bird2016},
which matches them against 
the overall amounts pledged
by the contributing funding agencies,
the procurement and operation 
of hardware resources 
are largely delegated to the individual sites,
resulting in a very diverse computing landscape.

The compute power provided by WLCG, in particular, 
comes from a variety of CPUs distributed worldwide,
where the specific hardware deployed at one site
can be quite different from that at another site,
both in terms of cost and of computing performance.
A~common unit of measurement is therefore needed 
to quantify the experiment needs and 
the resources provided by the sites 
in a given year,
and to allow review boards 
to compare these to the~amounts that were actually 
used~\cite{bib:bird-oct2019,bib:crsg-oct2019}.
A good evaluation metric 
of a compute resource, 
in this context, is one that is highly correlated 
to its application throughput, 
i.e. to the amount of useful ``work'' 
(e.g. the number of events processed by a HEP application)
that the compute resource can do per unit time: 
this is the typical use case 
of a CPU benchmark~\cite{bib:dongarra}.
Since 2009, in particular,
HEP-SPEC06 (HS06) has been 
the standard CPU benchmark for all of LHC computing.
The total integrated power of WLCG sites 
in 2017~\cite{bib:hsf2020},
for instance, was more 
than 5M HS06: 
taking into account that 
the typical worker nodes
deployed in WLCG have an HS06 score 
of around 10 per CPU core~\cite{bib:bwg},
this means that LHC computing in 2017
was supported by approximately 500k CPU cores.

While their main motivation 
is the overall accounting of resources,
both on a yearly basis
and in the planning of long term projects,
HS06 and other CPU benchmarks 
have many other applications in LHC computing.
Individual computing sites 
use HS06 for their procurement, 
to buy the CPU resources providing the 
amount of HS06 pledged to the HEP experiments
for the lowest financial cost,
also taking into account electrical power efficiency
measured in HS06 per Watt.
The experiments also use CPU benchmarks 
for scheduling~and managing their jobs
on distributed compute resources,
to predict the processing time required 
to complete a given application workload
and optimize its placement or smooth termination 
on batch queues~\cite{bib:philippe} 
and preemptible cloud resources.
CPU benchmarks may also be useful
in software optimizations,
to compare 
an application's performance
to the theoretical compute power 
of the machine where it is run,
or to the reference performance of another~application.

HS06 has served
the needs of WLCG for over 10 years, 
while many things have changed. 
On modern hardware,
users have reported scaling deviations 
up to 20\%~\cite{bib:philippe} 
from the performance predicted by HS06.
It is now clear that 
HS06 should be replaced by 
a new CPU benchmark.
In the following,
the motivations of the choice to develop
a new ``HEP-Benchmarks'' suite
and its implementation
are described.
After reviewing 
in Sec.~\ref{sec:hist}
the evolution of CPU benchmarks in HEP 
up until HS06,
Sec.~\ref{sec:hs06} 
describes the limitations of HS06
and the reasons why 
the new suite
is based on the containerization of 
LHC experiment software workloads.
Section~\ref{sec:bmk} summarises
the design, implementation choices and status
of HEP-Benchmarks,
while Section~\ref{sec:concl}
reports on its outlook,
a few months after the CHEP2019 conference.

\section{HEP CPU benchmarks: from the CERN Unit to HS06}
\label{sec:hist}

Computing architectures and software paradigms, 
in HEP and outside of it,
have significantly evolved over time 
and will keep on changing. 
This implies that CPU benchmarks, 
and more specifically those used for HEP computing, 
also need to evolve 
to keep up with these changes.

\paragraph{CERN Units}
The 1992 paper by Eric McIntosh~\cite{bib:mcintosh}
is an essential read 
to understand the beginnings
of CPU benchmarking in HEP
and its later evolution.
In the 1980's, 
the default benchmark 
was the ``CERN Unit'',
whose score was derived from a small set
of typical FORTRAN66 programs 
used in HEP at the time 
for event simulation and reconstruction.
This was used, for instance,
to grant CPU quotas to all users of CERN central systems.
In the early 1990's,
the definition of a CERN Unit was updated, 
as the task of running the old benchmark
on newer machines
turned out to be impossible:
the set of programs
was thus reviewed,
to make the benchmark more portable
and more representative of the then current HEP workloads
and of FORTRAN77.
It was already clear, however,
that HEP benchmarks should further evolve,
for instance 
to take into account 
a more widespread use 
of FORTRAN90,
of double-precision arithmetics on 32-bit architectures,
and possibly of vectorisation
and of parallel processing. 

Largely speaking, CPU benchmarks
can be grouped into three categories~\cite{bib:dixit}:
kernel, synthetic and application.
Kernel benchmarks
are based on libraries and small code fragments
that often account for the majority of CPU time 
in user applications:
an example is the LINPACK benchmark~\cite{bib:linpack},
based on the LINPACK matrix algebra package.
Synthetic benchmarks 
are custom-built to include 
a mix of low-level instructions,
e.g. floating-point or integer operations,
resembling that found 
in user applications.
Two examples 
are the Whetstone~\cite{bib:whet}
and Dhrystone~\cite{bib:dhry} benchmarks.
Application benchmarks are 
based on actual user applications.
From a user's perspective,
benchmarking a machine based on 
the user's own application
is clearly the best option, although 
in practice this is often impractical.

The CERN Unit had 
been designed
to be based, as much as possible, 
on real HEP applications,
rather than on kernel or synthetic benchmarks.
McIntosh made this clear in his 1992 paper,
where, however, he also commented that new approaches
may be needed for the future, 
as at the time
he considered it virtually impossible
to capture 
a modern event processing program
involving over 100k lines of code,
several external libraries
and one~or more databases or data sets.
For reference, 
the CERN Unit was shipped 
as a tarball that required less
than 50~MB of disk space in total,
for unzipping, building and executing all included programs,
using the compiler and operating system 
found on the machine to be benchmarked.

\paragraph{SPEC CPU benchmarks and SI2K (SPEC CINT2000)}
In his review of existing benchmarks outside HEP,
one option that McIntosh mentioned
as being perhaps the most useful for HEP
was the SPEC benchmark suite.
After the CERN Unit, indeed,
all of the default CPU benchmarks used in HEP have 
been based on the SPEC benchmark suite,
up until today.
SPEC (Standard Performance 
Evaluation Corporation)~\cite{bib:spec},
founded in 1988,
is a nonprofit corporation formed 
to establish, maintain and endorse standardized benchmarks 
of computing systems. 
SPEC distributes two different categories 
of CPU benchmark suites,
focusing on integer and floating-point operations. 
In the HEP world,
several versions of 
the SPEC CPU integer benchmark suite
have been used 
since the early 1990's,
starting with SPEC 
CPU92~\cite{bib:michelotto-paper}. 
In particular,
CINT2000 (the integer component of SPEC CPU2000),
known informally as SI2K,
was the CPU benchmark used in 2005
by the four LHC collaborations 
for their Computing Technical Design 
Reports~\cite{bib:alice-ctdr2005,bib:atlas-ctdr2005,bib:cms-ctdr2005,bib:lhcb-ctdr2005}.

Since about 2005, however,
many presentations at HEPiX conferences
pointed out a growing discrepancy between
the performances of HEP applications
and those~predicted~from the SI2K scores
of the systems where these applications were run.
In 2006, a HEPiX Benchmarking Working Group (BWG)
was set up specifically
with the task of identifying
the appropriate successor of SI2K.
In 2009, 
the BWG suggested~\cite{bib:michelotto-paper} 
to adopt a new HEP-specific benchmark, 
HEP-SPEC06 (HS06),
based on the then latest 
SPEC suite, CPU2006~\cite{bib:cpu2006}.

\paragraph{HEP-SPEC2006: a HEP-specific version 
of the SPEC CPU2006 C++ benchmark suite}
HS06 is based 
on a subset of SPEC CPU2006 
including the seven benchmarks written in C++~\cite{bib:wong},
three from the integer suite
and four from the floating-point suite.
In line with the general approach~\cite{bib:dixit} followed 
in SPEC CPU suites,
these seven programs 
are not kernel or synthetic benchmarks, 
but represent instead real applications,
mostly from scientific domains,
although not from the HEP domain.
HS06 differs from the SPEC CPU2006 C++ suite
in that it includes a few HEP-specific tunings:
for instance, 
the programs 
must be built using gcc 
in 32-bit mode also on 64-bit architectures,
and with other well defined compiler options~\cite{bib:bwg},
and they
must also be executed in a specific configuration
on the machine to be benchmarked,
as if the available processor cores 
were all configured as independent
single-core batch slots to run several 
single-process applications 
in parallel.

HS06 was identified as valid successor of SI2K
by the HEPiX BWG  
for essentially two reasons~\cite{bib:michelotto-paper}.
First, the HS06 score was found
to be highly correlated to throughput
on a large number of diverse machines
in a test ``lxbench'' cluster,
for each of many typical HEP applications.
The test machines were
typical WLCG worker nodes,
all based on x86 \mbox{architectures},
but including single-core and multi-core
CPUs with different speeds and from different vendors,
and with a diverse range of cache and RAM sizes. 
The test applications
covered four main HEP use cases~\cite{bib:benelli},
generation (GEN), simulation (SIM),
digitization (DIGI) and reconstruction (RECO),
including programs contributed 
by all four LHC experiments.
The~second reason for choosing HS06 
was that
its CPU usage pattern,
as measured from CPU hardware counters 
using perfmon~\cite{bib:eranian,bib:hirstius,bib:nowak},
was found to be quite similar to that 
observed on the CERN batch system 
used by the LHC experiments 
(in particular, the fraction of floating point
operations was around 10\% in both cases).
The memory footprint of the SPEC CPU2006 tests in HS06,
around 1~GB, was also comparable 
to that of typical HEP applications, requiring up to~2~GB
(while the memory footprint of the older SI2K benchmark 
was only 200~MB).

\section{The issues with HS06 and the choice of a new benchmark}
\label{sec:hs06}

In summary, 
in 2009 HS06 was chosen because,
while it is~based on a set of C++ applications
from domains other than HEP,
HS06 had been found 
to be sufficiently representative 
of HEP's own typical applications,
both in terms of throughput and of CPU usage patterns.
The problem today is that,
since a few years, it has become 
clear~\mbox{\cite{bib:chep2018,bib:acat2019,bib:trident}} 
\mbox{that this is no~longer the case}.

To start with,
the throughputs of HEP applications, 
mainly of ALICE and LHCb~\cite{bib:philippe},
have been reported to deviate up to 20\%
on some systems
from those predicted by HS06.
In addition,
important differences are now observed between
the CPU usage patterns of HS06 and HEP applications,
as measured from performance counters 
using Trident~\cite{bib:trident}
(a tool based on libpfm~\cite{bib:eranian} from perfmon):
in particular, 
with respect to the HS06 benchmarks,
HEP workloads have a lower
instructions-per-cycle (IPC) ratio
and may differ by 20\% or more
in the percentages of execution slots spent
in the four categories
suggested by Top-Down analysis~\cite{bib:yasin}
(retiring i.e. successful,
front-end bound, back-end bound 
and bad speculation).

More generally,
HS06 benchmarks
are no longer representative 
of WLCG software and computing today:
memory footprints have increased 
to 2 GB or more~\cite{bib:elms} per core;
64-bit builds have replaced 32-bit builds;
multi-threaded, multi-process 
and vectorized software solutions
are becoming more common;
and the hardware landscape 
is also more and more heterogeneous,
with the emergence of non-x86 architectures
such as ARM, Power9 and GPUs,
especially at HPC centers.

In addition, 
SPEC CPU2006,
on which HS06 is based,
was retired in 2018,
after the release 
of a newer SPEC CPU2017 benchmark suite.
An extensive analysis~\cite{bib:chep2018,bib:acat2019}
of this new suite by the HEPiX BWG,
however, pointed out that SPEC CPU2017 
is affected by the same problems as HS06.
In particular, SPEC CPU2017 scores
were found to have a high correlation 
to HS06 scores,
and hence 
still an unsatisfying correlation
to HEP workloads;
also, the CPU usage patterns 
of SPEC CPU2017,
as measured by Trident,
were found to be similar to those of HS06,
and quite different from those of HEP workloads.

\paragraph{The HEP-Benchmarks suite: 
using containerized HEP workloads 
as CPU benchmarks}
The solution to the issues described above is,
in theory, quite simple.
Rather than testing 
real application benchmarks
from domains other than HEP
(like those in SPEC CPU2006 and CPU2017)
or kernel or synthetic benchmarks,
and looking for the benchmarks
whose score has the highest correlation
to the throughputs of typical HEP workloads,
and whose CPU usage patterns
look most similar to those of HEP workloads,
the ``obvious'' approach to follow
is to build a benchmark suite
including precisely those typical HEP workloads.

{\it By construction}, in fact,
a benchmark based on a HEP application
is guaranteed to give 
a score and a CPU usage pattern
that are the most representative
of that application.
This is precisely the approach
followed in 
the new HEP-Benchmarks~\cite{bib:hep-bmk} suite,
which we are building within the HEPiX BWG
to make it the successor of HS06,
as described in the next section.
The central package 
of this project,
hep-workloads,
is a collection of 
workloads from the four LHC experiments,
covering all of the
GEN, SIM, DIGI and RECO use cases.

In retrospective, this is the same approach 
on which the CERN Unit was based.
As~discussed in the previous section,
the CERN Unit 
was eventually discontinued 
because in the early 1990's
it seemed no longer possible
to capture a complex HEP application
with all~of its software and data dependencies.
A problem which seemed impossible to solve 
30 years ago,
however, can be much more easily addressed
using the technologies available today. 
The reason why it is now possible 
to encapsulate HEP workloads
in the hep-workloads package,
in particular,
is the availability 
of two enabling technologies:
first and foremost,
Linux containers~\cite{bib:docker},
which allow the packaging and distribution of
HEP applications
with all of their dependencies,
including the full O/S;
and, in addition,
the cvmfs shrinkwrap utility~\cite{bib:cvmfs-sw,bib:cvmfs-sw2},
which makes it possible to 
selectively capture 
which specific software and data files
are needed to execute a HEP workload
in a portable and reproducible way,
out of the much larger 
LHC experiment software installations
on the cvmfs (CernVM-FS) filesystem~\cite{bib:cvmfs}.

\section{HEP-Benchmarks: a new CPU benchmark suite for HEP}
\label{sec:bmk}

The development of 
what has now become
the HEP-Benchmarks project 
started in \mbox{mid-2017} 
as a proof-of-concept study
by one member of the HEPiX BWG,
using the ATLAS kit-validation (KV)~\cite{bib:kv}
and a CMS workload as first examples. 
This work took off on a larger scale
towards the end of 2018,
when many more collaborators
from the BWG and the four LHC experiments joined the effort.
The project is maintained 
on the gitlab infrastructure at CERN~\cite{bib:hep-bmk},
which is also used for Continuous Integration (CI) 
builds and tests, 
for issue tracking and for documentation.
HEP-Benchmarks includes three main components,
which are mapped to separate gitlab repositories
and are described in the following subsections.

\paragraph{The hep-workloads package: HEP reference workloads} 
The hep-workloads package
is the core of the HEP-Benchmarks suite.
It contains all 
the code and infrastructure,
both common and workload-specific,
to build a standalone 
container for each of the 
HEP software workloads it includes.
Individual images
are built, tested and versioned in the package's gitlab CI and 
are then distributed 
via its container registry~\cite{bib:hep-wl}.
Images are built as Docker containers~\cite{bib:docker},
but~they can also be executed 
via Singularity~\cite{bib:sing}.

A single command is enough to download
a specific benchmark and execute it
using~the embedded pre-compiled 
libraries and binaries.
A benchmark summary file in json format and more detailed logs 
are stored in a results directory.
For instance,
to download the latest version of the LHCb GEN/SIM image, 
execute it using either Docker or Singularity,
and store results in the host /tmp directory,
it is enough to run
one of 
the two following~commands:
\begin{lstlisting}
  docker run -v /tmp:/results \ gitlab-registry.cern.ch/hep-benchmarks/hep-workloads/lhcb-gen-sim-bmk:latest
  singularity run -B /tmp:/results \ gitlab-registry.cern.ch/hep-benchmarks/hep-workloads/lhcb-gen-sim-bmk:latest
\end{lstlisting}
The main result 
in the json file
is the benchmark score for the given workload,
measured as an absolute 
throughput of number of events processed per wall-clock time.
This~is~essentially derived 
from the total wall-clock time to complete
the processing of a predefined number of events.
The json file also contains 
all relevant metadata 
about how the benchmark was run,
as well as more detailed results,
including memory and CPU time usage.

Some of the HEP workloads, 
like ALICE and LHCb GEN/SIM,
are single-process~(SP) and single-threaded (ST);
others use parallelism
to save memory on multi-core CPUs, 
via multi-threading (MT) techniques
like CMS RECO~\cite{bib:cmsswmt},
or via multi-process (MP) techniques
involving forking and copy-on-write,
like ATLAS RECO~\cite{bib:athenamp}.
For MT/MP applications,
the number of threads/processes 
is fixed to that used 
by the experiment
in production,
and every image is executed 
in such a way as to fill 
all available logical cores 
on the machine that is benchmarked,
by launching an appropriate number 
of identical copies of each application.
When 
more than one copy is executed, 
the benchmark score is 
the sum of their throughputs.
The number of logical cores,
derived from the 
\lstinline[basicstyle={\small\ttfamily}]{nproc} command,
is equal to the number of physical cores
for machines configured with hyper-threading disabled,
but it is higher if
this is enabled.
On a machine with 16 physical cores 
and 2x hyper-threading, for instance,
by default
32 copies of the ST/SP LHCb GEN/SIM 
and 8 copies of the 4xMT CMS RECO benchmarks
are executed. 
All of these parameters are, in any case,
configurable.

\newcommand{\cvmfs}{\lstinline[basicstyle={\small\ttfamily}]{/cvmfs}}
The design of the hep-workloads package
relies on the fact 
that all four LHC experiments
install and distribute 
their pre-compiled software libraries
using the cvmfs file system.
To~add a new workload, 
experiment experts just need
to prepare an orchestrator script,
which sets the 
runtime environment,
runs one or more copies of the application, and finally 
parses the output logs 
to determine the event throughput,
which is used as benchmark score.
A common benchmark driver
harmonises the control flow
and error checking in all workloads,
making it easier to debug them.
The build procedure in the gitlab CI 
includes the following four steps.
\begin{enumerate}
\item 
\vspace*{-1mm}
First, 
an interim Docker container is built,
where \cvmfs\ is, as usual,
a directory managed by the network-aware cvmfs service~\cite{bib:cvmfs}, 
which is able to retrieve missing files via http.
In addition, the cvmfs shrinkwrap~\cite{bib:cvmfs-sw,bib:cvmfs-sw2}
tracing mechanism is enabled.
\item 
\vspace*{-1mm}
One copy of the workload application 
from that interim image is then executed:
this generates a shrinkwrap trace file specifying 
which files were accessed from \cvmfs. 
\item
\vspace*{-1mm}
The final standalone Docker container is built,
where \cvmfs\ is a local folder,
including all files identified
by tracing during the previous step.
This includes 
all relevant experiment libraries and executables,
pre-compiled for the O/S chosen for this image.
\item 
\vspace*{-1mm}
This final container
is tested, by running the workload 
using both Docker and Singularity.
If tests succeed, the image is pushed
to the gitlab registry~\cite{bib:hep-wl}.
\end{enumerate}
A key element of this approach
is reproducibility: 
repeated runs of each workload 
are meant to always process the same events
and produce the same results.
This is essential for resource benchmarking,
to ensure that timing measurements 
on two different nodes correspond to 
the same computational load.
It is also important during the build process,
where tests of the final container may fail
if they need different \cvmfs\ files
from those identified while tracing the interim container.
Strict reproducibility can be guaranteed 
for ST (including MP) workloads,
but not for the CMS MT workloads, 
where the sharing of processing
across different software threads
may lead to small differences in execution paths;
however, this is not considered an issue for benchmarking,
and no errors have been observed during the build process either.

Workload images vary in size 
between 500 MB (ATLAS GEN) and 4 GB (CMS DIGI).
GEN containers are generally the smallest
because event generation is 
CPU intensive with almost no input data,
while DIGI and RECO images are much larger
as event digitisation and reconstruction 
are more I/O intensive 
and large reference data files must be shipped 
within the workload containers.
Docker images are internally made up of layers
(and this structure is maintained 
when they are converted to Singularity).
Taking into account that 
bug fixing and feature improvements
have often led to a rapid development cycle,
the hep-workloads CI has been optimized to stack these layers
in the order which makes them as cacheable as possible.
The bottom layers contain what changes least often, 
like the O/S and data files, 
while the higher layers include experiment software
and common and workload-specific scripts.

\paragraph{The hep-score package: 
a new CPU benchmark for HEP}

The aim of the hep-score package
is to combine the benchmark scores 
derived from the individual HEP workloads
into a single number, a ``HEPscore''.
The package is highly configurable,
allowing the definition of a combined HEPscore
from any combination of specific versions 
of individual workloads,
with specific MT/MP settings.
The numbers of events to process can also be tuned,
to choose the appropriate compromise
between benchmark precision and execution time.
The prototypes that are currently being developed, 
for instance, derive 
a combined score
from a geometric mean of the throughputs of 
ATLAS (GEN,~SIM and DIGI/RECO), CMS (GEN/SIM, DIGI and RECO)
and LHCb (GEN/SIM) benchmarks,
and take between 6 and 16 hours to complete.
This is similar to what was done for HS06,
whose combined score was derived 
from the geometric mean of the~7 
individual SPEC CPU2006 C++ benchmarks.
A normalization factor can also be~added
for each individual benchmark,
to redefine its relative score on a machine 
as a ratio, i.e. as the throughput on that machine divided by 
the throughput on the reference machine.
Unlike absolute throughputs,
which can take a priori any value
and have the dimensions of events processed per second,
these relative scores are quite practical
because they are adimensional numbers.
In particular, 
the relative scores of individual benchmarks,
and a fortiori the combined relative score 
defined as the geometric mean 
of a subset of these scores, are all 
equal to~1 on the reference machine.

Having said that,
it should be stressed that it is impossible 
to characterize a computer system's performance
by a single metric.
This concept was very well expressed by Kaivalya Dixit,
long-time president of the SPEC corporation,
who even warned 
about ``the danger and the folly''~\cite{bib:dixit}
of relying on either a single performance number or a single benchmark.
There are, however, many use cases
where a single number is needed,
and the accounting of WLCG resources
is presently one of them.
This is not a technical issue: it is a policy issue, 
and its
discussion is beyond the scope of this paper.
On a technical level,
our design of the hep-score package 
takes this into account 
by allowing the definition 
of a highly configurable combined score,
but also by the fact that 
the detailed scores of individual workloads
are also stored 
in the report of any HEPscore execution.
This is very important because it provides a mechanism
to analyse a posteriori 
the performance of
individual~HEP~workloads.

\paragraph{The hep-benchmark-suite package: 
a toolkit for benchmark execution and result collection}

The hep-benchmark-suite package, finally,
is a toolkit to coordinate the execution of
several benchmarks,
including not only HEPscore,
but also HS06, SPEC CPU2017, KV and others.
Results are collected in a global json document
that can then be uploaded to a database.

This package is used
for what are currently the main
activities of our team,
to demonstrate the readiness
of HEP-Benchmarks as a replacement of HS06:
first, for testing that individual HEP workloads
are reliable and give stable results 
(typically, within 5\% or better)
on repeated run on the same system;
second, for
the study of their correlations
to one another, and to HS06 and other benchmarks.
To this end,
a wide range of x86 worker nodes
has been collected,
similar to the lxbench cluster
that had been used in the initial comparisons 
of HS06 and SI2K.

\section{Outlook: GPUs and non-x86 CPU architectures}
\label{sec:concl}

To date,
WLCG pledged compute resources
have essentially consisted only of x86 CPUs.
This processor architecture has therefore been the 
main focus of developments in the HEP-Benchmarks project
so far.
The design of its components, however, 
and specifically that of the core hep-workloads package,
is quite general and can be easily extended
to non-x86 architectures such as ARM or Power9,
and even to other compute resources such as GPUs,
which are becoming important for WLCG 
because of their widespread adoption 
in the latest 
\mbox{supercomputers} at HPC centers.
By and large, 
the large scale GEN, SIM, DIGI, RECO production workloads 
of the four LHC experiments are not yet ready~\cite{bib:avgpu}
to be moved from traditional WLCG x86 resources to GPUs, but 
we should be ready to benchmark these resources when this happens.
Within the HEP-Benchmarks project,
a new hep-workloads-GPU package has therefore been added,
to prototype the benchmarking of GPU workloads,
including a software workload from the LHC accelerator domain,
SixTrack~\cite{bib:sixtrack}.
Work is also in progress to integrate
a prototype of the CMS RECO workload 
on heterogeneous resources~\cite{bib:patatrack}.

\end{document}